\pdfoutput=1

\documentclass[pdftex,twocolumn,10pt,letterpaper]{article}

\usepackage{graphicx, multirow, times}
\usepackage{url, amsfonts, amsmath, verbatim, mathtools, soul}

\usepackage[small]{title sec}
\usepackage[status=draft]{fixme}
\usepackage[printwatermark]{xwatermark}

\usepackage{epsfig}
\usepackage{amssymb}
\usepackage{authblk}
\usepackage{grffile} 

\newcommand{\xref}[1]{\S\ref{#1}}
\newcommand{\panictwo}[1]{\vspace{-#1 plus 2pt minus 2pt}}
\newcommand{\ncaption}[1]{
  \panictwo{4pt}
  \renewcommand{\baselinestretch}{0.85}
  \caption{\small #1}
  \panictwo{2pt}
  \renewcommand{\baselinestretch}{1}
}

\usepackage[pagebackref=false,breaklinks=true,colorlinks,bookmarks=false]{hyperref}

\begin{document}

\title{\Large Occupy the Cloud: Distributed Computing for the 99\%}


\author{Eric Jonas, Qifan Pu, Shivaram Venkataraman, Ion Stoica, Benjamin Recht\\University of California, Berkeley\\Submission Type: Vision}

\date{}

\maketitle
\begin{abstract}
  Distributed computing remains inaccessible to a large number of users, in spite of many open source platforms and extensive commercial offerings. While distributed computation frameworks have moved beyond a simple
  map-reduce model, many users are still left to struggle with complex cluster management and configuration tools, even for running simple embarrassingly parallel jobs. 
  We argue that stateless functions represent a viable platform for these users, eliminating cluster management overhead, fulfilling the promise of elasticity. Furthermore, using our prototype implementation, PyWren, we show that this model is general enough to implement a number of distributed computing models, such as BSP, efficiently.
Extrapolating from recent trends in network bandwidth and the advent of disaggregated storage, we suggest that stateless functions are a natural fit for data processing in future computing environments. 
\end{abstract}





\section{Introduction}
Despite a decade of availability, the twin promises of scale and elasticity~\cite{armbrust2010view} remain out of reach for a large number of cloud computing users.
Academic and commercially-successful platforms (Apache Hadoop, Apache Spark) with tremendous corporate backing (Amazon, 
Microsoft, Google) still present high barriers to entry for the average data scientist or scientific computing user. In fact, taking advantage of elasticity remains challenging for even sophisticated users, as the majority of these frameworks were designed to first target on-premise installations at large scale. On commercial cloud platforms, a novice user confronts a dizzying array of potential decisions: one must ahead of time decide on instance type, cluster size, pricing model, programming model, and task granularity.

Such challenges are particularly surprising considering that the vast number of data analytic and scientific computing workloads remain embarrassingly parallel.
Hyperparameter tuning for machine learning, Monte Carlo simulation for computational physics, and featurization for data science all fit well into a traditional map-reduce framework. Yet even at UC Berkeley, we have found via informal surveys that the majority of machine learning graduate students have never written a cluster computing job due to complexity of setting up cloud platforms.

In this paper we argue that a \emph{serverless} execution model with \emph{stateless} functions can enable radically-simpler, fundamentally elastic, and more user-friendly distributed data processing systems. In this model, we have one simple primitive: users submit functions that are executed in a remote container; the functions are stateless as all the state for the function, including input, output is accessed from shared remote storage. Surprisingly, we find that the performance degradation from using such an approach is negligible for many workloads and thus, our simple primitive is in fact general enough to implement a number of higher-level data processing abstractions, including MapReduce and parameter servers.


Recently cloud providers (e.g., AWS Lambda, Google Cloud Functions) and open source projects (e.g., OpenLambda~\cite{hendrickson2016serverless}, OpenWhisk~\cite{openwhisk}) have developed infrastructure to run event-driven, stateless functions as micro-services. In this model, a function is deployed once and is invoked repeatedly whenever new inputs arrive and elastically scales with input size. Our key insight is that we can dynamically inject code into these functions, which combined with remote storage, allows us to build a data processing system that inherits the elasticity of the serverless model while addressing the simplicity for end users.



We describe a prototype system, PyWren\footnote{PyWren is available at https://pywren.io}, developed in Python with AWS Lambda. By employing only stateless functions, PyWren helps users avoid the significant developer and management overhead that has until now been a necessary prerequisite. The complexity of state management can instead be captured by a global scheduler and fast remote storage. With PyWren, we seek to understand the trade-offs of using stateless functions for large scale data analytics and specifically what is the impact of solely using remote storage for inputs and outputs. We find that we can achieve around 30-40 MB/s write and read performance per core to a remote bulk object store (S3), matching the per-core performance of a single local SSD on typical EC2 nodes. Further we find that this scales to 60-80 GB/s to S3 across 2800 simultaneous functions, showing that existing remote storage systems may not be a significant bottleneck.

Using this as a building block we implement image processing pipelines where we extract per-image features during a map phase via unmodified Python code. We also show how we can implement BSP-style applications on PyWren and that a word count job on 83M items is only 17\% slower than PySpark running on dedicated servers. Shuffle-intensive workloads are also feasible as we show PyWren can sort 1TB data in 3.4 minutes. However, we do identify storage throughput as a major bottleneck for larger shuffles. Finally we discuss how parameter servers, a common construct in distributed ML~\cite{li2014scaling} can be used with this model. We conclude the paper with some remaining systems challenges, including launch overhead, storage performance and scalable scheduling.






\section{Is the cloud usable?}
\label{sec:motivation}
Most software, especially in scientific and analytics applications, is not written by computer scientists~\cite{hettrick_2014_14809, momcheva2015software}, and it is many of these users who have been left out of the cloud revolution. 

The layers of abstraction present in distributed data processing platforms are complex and difficult to correctly configure. For example, PySpark, arguably one of the easier to use platforms, runs on top of Spark~\cite{zaharia2010spark} (written in Scala) which interoperates and is closely coupled with HDFS~\cite{shvachko2010hadoop} (written in Java), Yarn~\cite{yarn} (Java again), and the JVM. The JVM in turn is generally run on virtualized Linux servers. Merely negotiating the memory limit interplay between the JVM heap and the host operating system is an art form~\cite{fang2015interruptible,spark-gc-tuning,spark-gc-blog}. These systems often promote ``ease of use" 
by showing powerful functionality with a few lines of code, but this ease of use means little without mastering the configuration of the layers below.

In addition to the software configuration issues, cloud users are also immediately faced with tremendous planning and workload management before they even begin running a job. AWS offers 70 instances types across 14 geographical datacenters -- all with subtly different pricing. This complexity is such that recent research has focused on algorithmic optimization of workload trade-offs~\cite{herodotou2011starfish,venkataraman2016ernest}.  While several products such as Databricks and Qubole simplify cluster management,  the users still need to explicitly start and terminate clusters, and pick the number and type of instances.

Finally, the vast majority of scientific workloads could take advantage of dynamic market-based pricing of servers, such as AWS spot instances -- but computing spot instance pricing is challenging, and additionally most of the above-mentioned frameworks make it difficult to handle machine preemption. To avoid the risk of losing intermediate data, users must be careful to either regularly checkpoint their data or run the master and a certain number of workers on non-spot instances. This adds another layer of management complexity which makes elasticity hard to obtain in practice. 



\subsection{What users want}
Our proposal in this paper was motivated by a professor of computer graphics at UC Berkeley asking us ``Why is there no cloud button?" He outlined how his students simply wish they could easily ``push a button" and have their code -- existing, optimized, single-machine code -- running on the cloud. Thus, our fundamental goal here is to allow as many users as possible to take existing, legacy code and run it in parallel, exploiting elasticity. 
In an ideal world, users would simply be able to run their desired code across a large number of machines, bottlenecked only by serial performance. Executing 100 or 10000 five-minute jobs should take roughly five minutes, with minimal start-up and tear-down overhead.

Further, in our experience far more users are capable of writing reasonably-performant single-threaded code, using numerical linear algebra libraries (e.g., OpenBLAS, Intel's MKL), than writing complex distributed-systems code. Correspondingly the goal for these users is not to get the best parallel performance, but rather to get vastly better performance than available on their laptop or workstation while taking \emph{minimal development time}.


For compute-bound workloads, it is more useful to parallelize across functions rather than within each function; to say sweep over a wide range of parameters (such as machine learning hyperparameter optimization) or try a large number of random initial seeds (Monte Carlo simulations of physical systems). For these users, a simple function interface that captures sufficient local state, performs computation remotely, and returns the result is more than adequate. 
For data-bound workloads, a large number of users would be served by a simpler version of the existing map-reduce framework where outputs can be easily persisted on object storage. 

Thus, a number of compute-bound and data-bound workloads can be captured by having a simple abstraction that allows users to run arbitrary functions in the cloud without setting up and configuring servers/frameworks etc. We next discuss why such an abstraction is viable now and the components necessary for such a design.


\begin{figure}[!t]
    \centering
    \includegraphics[width=0.9\columnwidth]{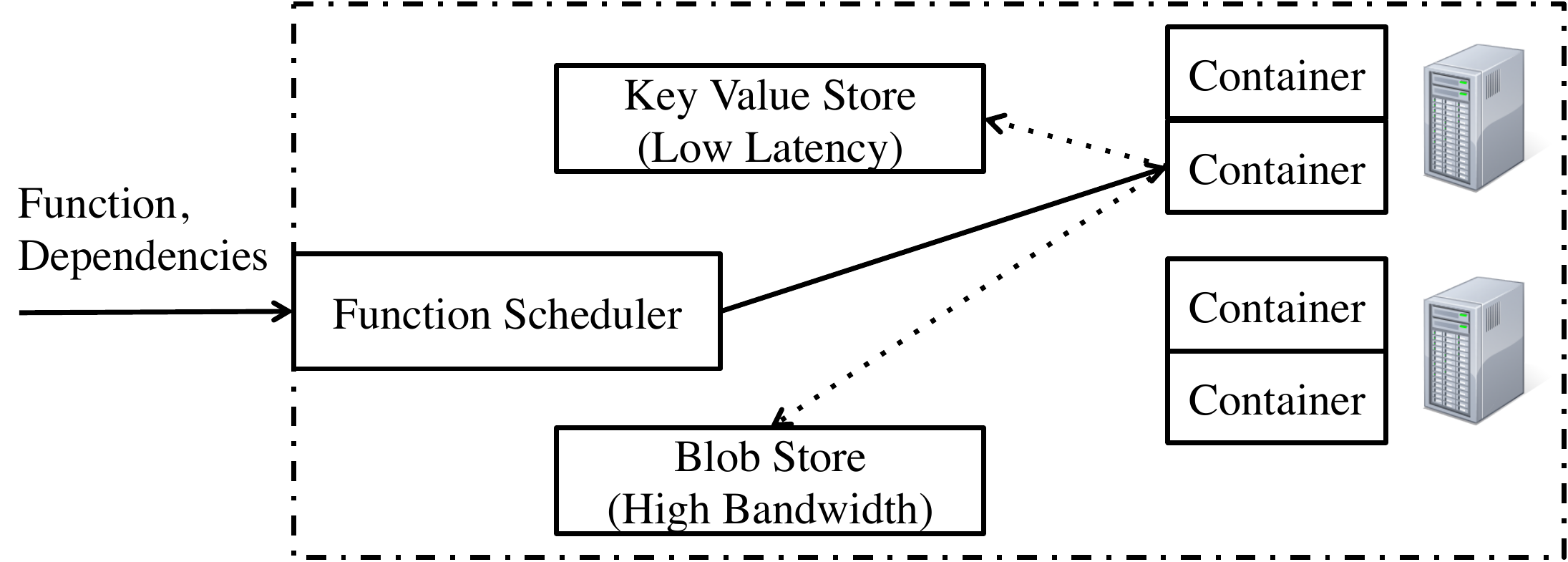}
    \ncaption{\bf System architecture for stateless functions}
    \label{fig:system-arch}
    \vspace{-0.2in}
\end{figure}

\section{A Modest Proposal}
Many of the problems with current cloud computing abstractions stem from the fact that they are designed for a server-oriented resource model. Having servers as the unit of abstraction ties together multiple resources like memory, CPU and network bandwidth. Further servers are also often long running and hence require DevOps support for maintenance. 
Our proposal is to instead use a serverless architecture with \emph{stateless functions} as the unifying abstraction for data processing. Using stateless functions will simplify programming and deployment for end users.
In this section we present the high level components for designing data processing systems on a serverless architecture. While other proposals~\cite{aws-lambda-mapred} have looked at implementing data processing systems on serverless infrastructure, we propose a simple API that is tightly integrated with existing libraries and also study performance trade-offs of this approach by using our prototype implementation on a number of workloads.  


\subsection{Systems Components}
The main components necessary for executing stateless functions include a low overhead execution runtime, a fast scheduler and high performance remote storage as shown in Figure~\ref{fig:system-arch}. 
Users submit single-threaded functions to a global scheduler and while submitting the function they can also annotate the runtime dependencies required. 
Once the scheduler determines where a function is supposed to run, an appropriate container is created for the duration of execution. While the container maybe reused to improve performance none of the state created
by the function will be retained across invocations. Thus, in such a model all the inputs to functions and all output from functions need to be persisted on remote storage and we include client libraries to access both high-throughput and low latency shared storage systems.

\textbf{Fault Tolerance}: Stateless functions allow simple fault tolerance semantics. When a function fails, we restart it (at possibly a different location) and execute on the same input. We only need atomic writes to remote storage for tracking which functions have succeeded. Assuming that functions are idempotent we obtain similar fault tolerance guarantees as existing systems.

\textbf{Simplicity}: As evidenced by our discussion above, our architecture is very simple and only consists of the minimum infrastructure required for executing functions. We do not include any distributed data structures or dataflow primitives in our design. We believe that this simplicity is necessary in order to make simple workloads like embarrassingly parallel jobs easy to use. More complex abstractions like dataflow or BSP can be implemented on top and we discuss this in Section~\ref{sec:apps}. 
%

\begin{table}[!t]
\begin{center}
\begin{tabular}{c|c}
\hline
  Storage Medium & Write Speed (MB/s) \\
\hline
  SSD on \texttt{c3.8xlarge} & 208.73 \\
  SSD on \texttt{i2.8xlarge} & 460.36 \\
  4 SSDs on \texttt{i2.8xlarge} & 1768.04 \\
  S3 & 501.13 \\
\end{tabular}
\ncaption{\bf Comparison of single-machine write bandwidth to instance local SSD and remote storage in Amazon EC2. Remote storage is faster than single SSD on the standard \texttt{c3.8xlarge} instance and the storage-optimized \texttt{i2.8xlarge} instance.}
\label{tab:ssd_vs_s3}
\end{center}
\vspace{-0.8cm}
\end{table}

\textbf{Why now?} While the model described above is closely related to systems like Linda~\cite{carriero1998linda}, Celias~\cite{han2013large} and database trigger-based systems~\cite{power2010piccolo,peng2010large}, these systems have not been widely adopted. We believe that this model
is viable now given existing infrastructure and technology trends. While the developer has no control of where a stateless function runs (e.g., the developer cannot specify that a stateless function should run on the node storing the function's input), the benefits of colocating computation and data -- a major design goal for prior systems like Hadoop, Spark and Dryad -- have diminished.

Prior work has shown that hard disk locality does not provide significant performance benefits~\cite{diskirrelevant}. To see whether the recent datacenter migration from hard disks to SSDs has changed this conclusion, we benchmarked the I/O throughput of storing data on a local SSD of an AWS EC2 instance vs. storing data on S3. Our results, in Table~\ref{tab:ssd_vs_s3}, show that currently that writing to remote storage is faster than a single SSD but using multiple SSDs can yield better performance.
However, technology trends~\cite{han2013network,colin-trends,ethernet-standards} indicate that the gap between network bandwidth and storage I/O bandwidth is narrowing, and many recently published proposals for rack-scale computers feature disaggregated storage~\cite{hp-machine, asanovic2014firebox} and even disaggregated memory~\cite{gao2016network}. All these trends suggest diminishing performance benefits from colocating compute with data in the future.






\begin{figure*}[!t]
  \centering
  \begin{minipage}[t]{0.3\textwidth}
    \begin{center}
        \includegraphics[width=\columnwidth]{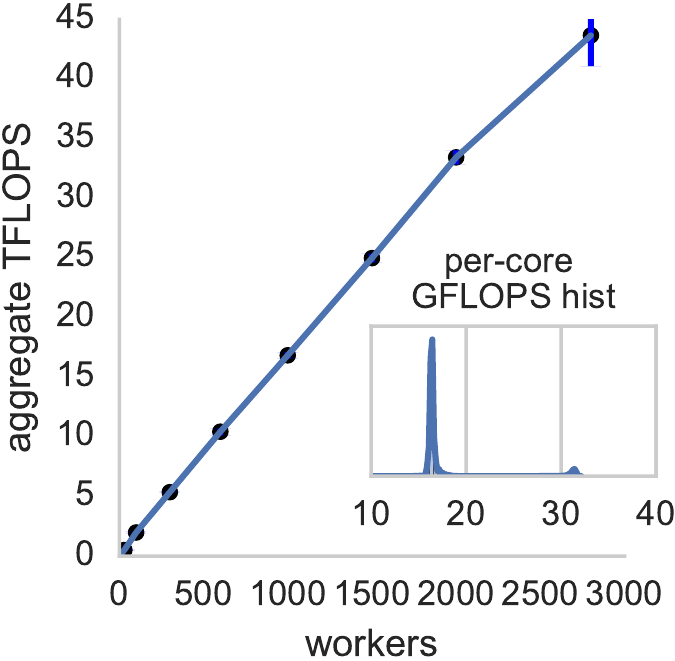}
    \end{center}
    \vspace{-0.2in}
    \ncaption{\bf Running a matrix multiplication benchmark inside each worker, we see a linear scalability of FLOPs across 3000 workers.}
    \label{fig:microbench:flops}
  \end{minipage}
  \hspace{.01\textwidth}
  \begin{minipage}[t]{0.3\textwidth}
    \begin{center}
        \includegraphics[width=\columnwidth]{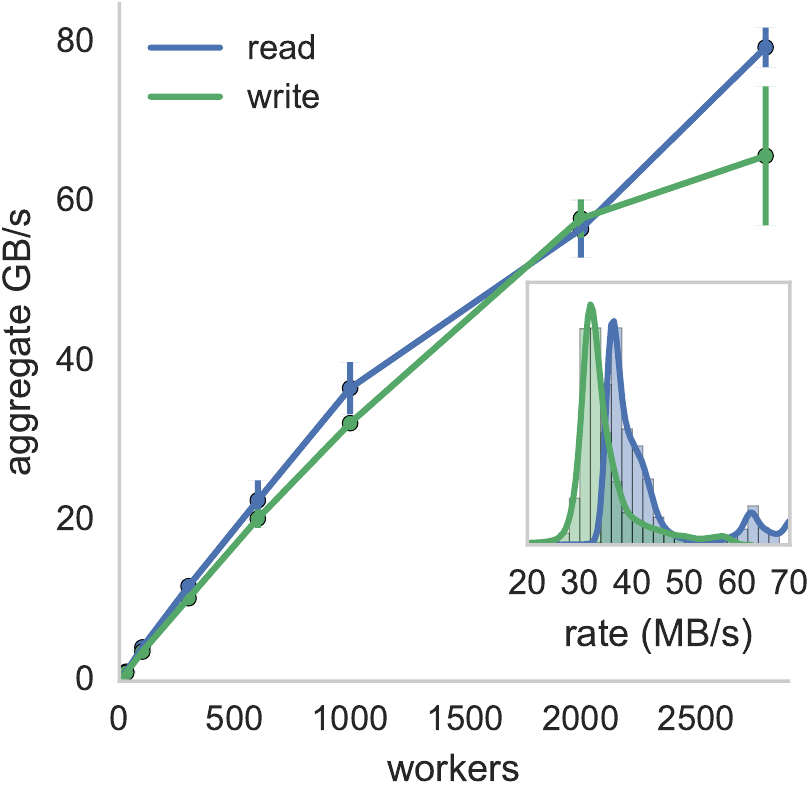}
    \end{center}
    \vspace{-0.2in}
    \ncaption{\bf Remote storage on S3 linearly scales with each worker getting around 30 MB/s bandwidth (inset histogram).}
    \label{fig:microbench:s3}
  \end{minipage}
  \hspace{.01\textwidth}
  \begin{minipage}[t]{0.3\textwidth}
    \begin{center}
        \includegraphics[width=\columnwidth]{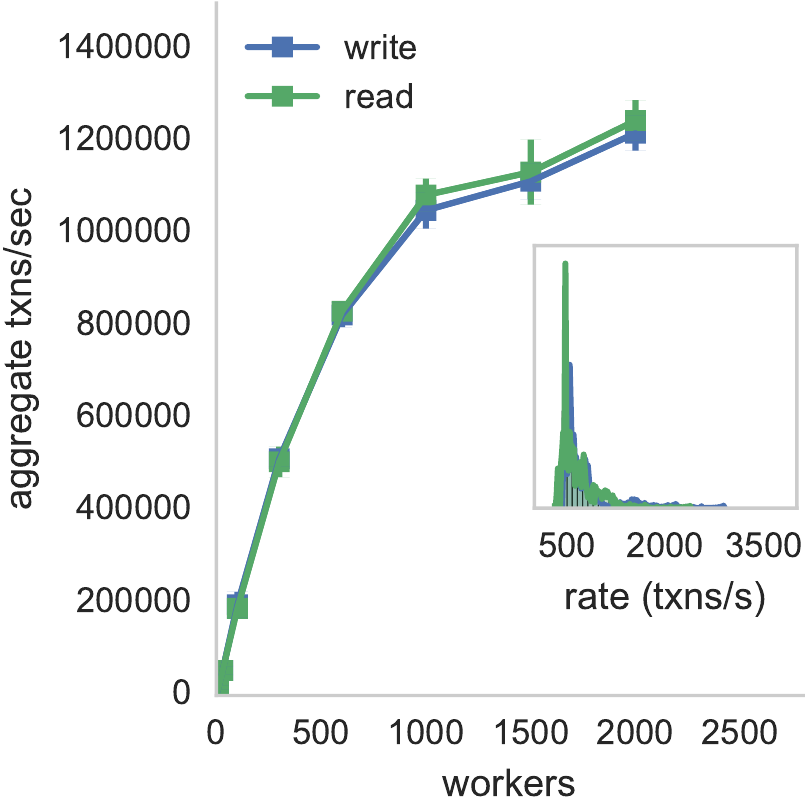}
    \end{center}
    \vspace{-0.2in}
    \ncaption{\bf Remote key-value operations to Redis scales up to 1000 workers. Each worker gets around 700 synchronous transactions/sec.}
    \label{fig:microbench:redis}
  \end{minipage}
  \vspace{-0.2in}
\end{figure*}

\subsection{PyWren: A Prototype}
\label{sec:pywren}
We developed PyWren\footnote{A wren is much smaller than a Condor} to rapidly evaluate these ideas, exposing a seamless map primitive from Python on top of AWS Lambda. While Lambda was designed to run event-driven microservices (such as resizing a single user-uploaded image) at scale, by dynamically extracting code from S3 we make each Lambda invocation run a different function. Currently AWS Lambda provides a very restricted containerized runtime with a maximum 300 seconds of execution time, 1.5 GB of RAM, 512 MB of local storage and no root access, but we believe these limits will be increased as AWS Lambda is used for more general purpose applications. 




PyWren serializes a Python function using \texttt{cloudpickle}~\cite{cloudpickle}, capturing all relevant information as well as most modules that are not present in the server runtime\footnote{While there are limitations in the serialization method (including an inability to transfer arbitrary python C extensions), we find this can be overcome using libraries from package managers such as Anaconda.}. This eliminates the majority of user overhead about deployment, packaging, and code versioning. We submit the serialized function along with each serialized datum by placing them into globally unique keys in S3, and then invoke a common Lambda function. On the server side, we invoke the relevant function on the relevant datum, both extracted from S3. The result of the function invocation is serialized and placed back into S3 at a pre-specified key, and job completion is signaled by the existence of this key. In this way, we are able to reuse one registered Lambda function to execute different user Python functions and mitigate the high latency for function registration, while  executing functions that exceed Lambda's code size limit. 

\textbf{Map for everyone}: As discussed in Section~\ref{sec:motivation}, many scientific and analytic workloads are embarrassingly parallel. The \texttt{map} primitive provided by PyWren makes addressing these use cases easy -- serializing all local state necessary for computation, transparently invoking functions remotely and returning when complete. Calling \texttt{map} launches as many stateless functions as there are elements in the list that one is mapping over.  An important aspect to note here is that this API mirrors the existing Python API for parallel processing and thus, unlike other serverless MapReduce frameworks~\cite{aws-lambda-mapred}, this integrates easily with existing libraries for data processing and visualization. 

\begin{figure}[t]
  \begin{minipage}[]{\columnwidth}
    \centering
    \includegraphics[width=0.95\columnwidth]{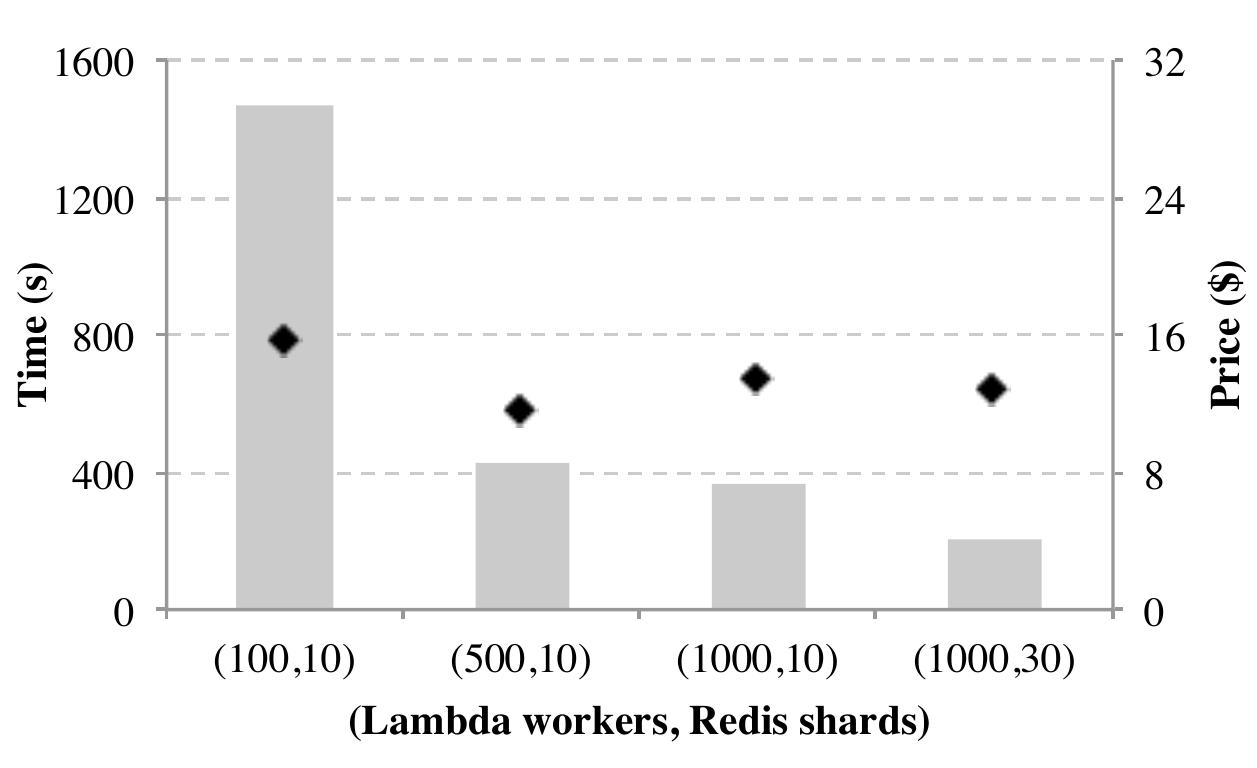}
    \vspace{-0.3cm}
    \ncaption{\bf Prorated cost and performance for running 1TB sort benchmark while varying the number of Lambda workers and Redis shards.}
    \label{fig:sort_cost}
    \vspace{0.1cm}
    \centering
    \includegraphics[width=0.95\columnwidth]{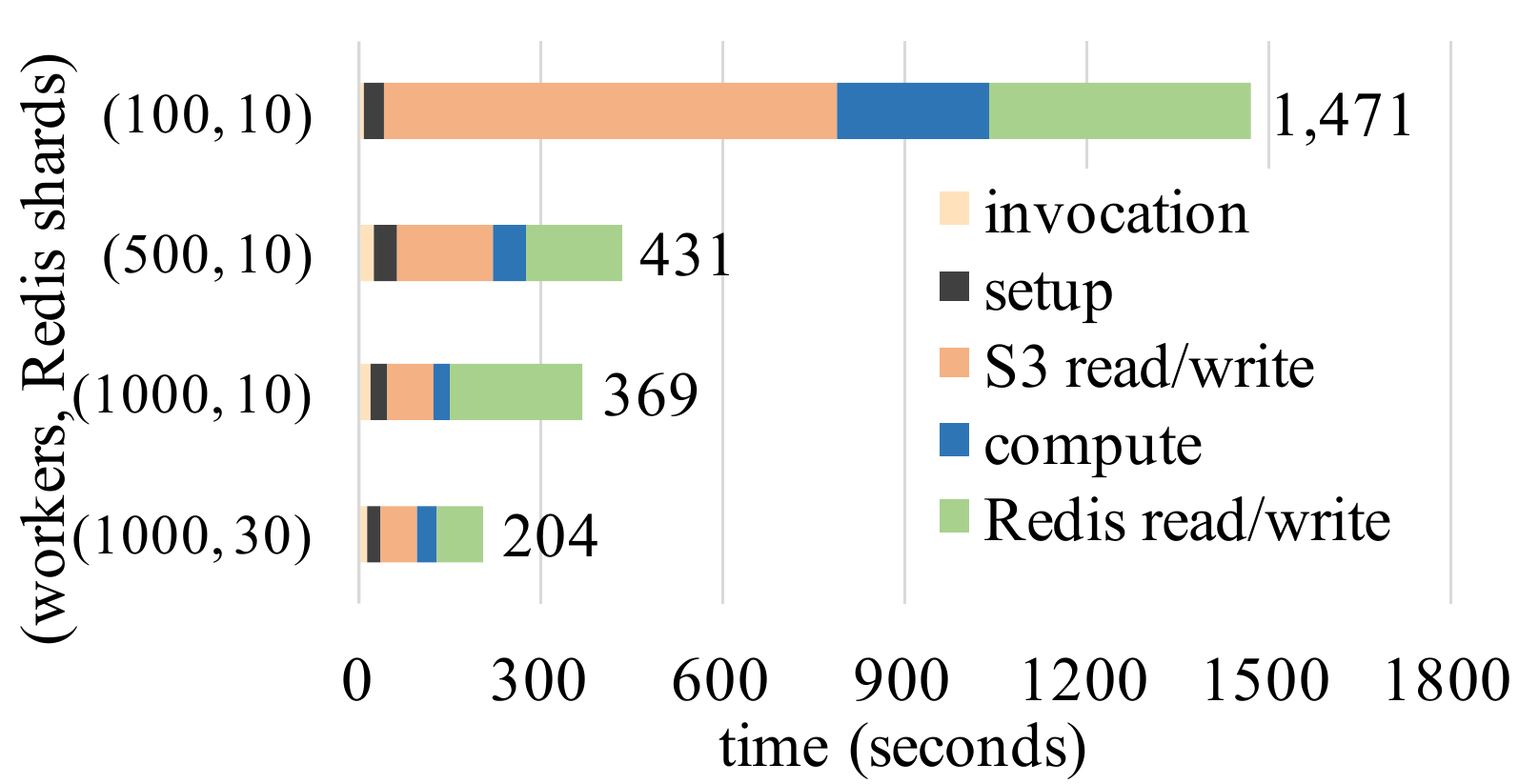}
    \vspace{-0.3cm}
    \ncaption{\bf Performance breakdown for sorting 1TB data by how task time is spent on average.}
    \label{fig:sort_time}
  \end{minipage}
  \label{fig:sort}
\vspace{-0.6cm}
\end{figure}

\textbf{Microbenchmarks}: Using PyWren we ran a number of benchmarks(Figures~\ref{fig:microbench:flops},\ref{fig:microbench:s3},\ref{fig:microbench:redis}) to determine the impact of solely using remote storage for IO, and how this scales with worker count. In terms of compute, we ran a matrix multiply kernel within each Lambda and find that we get ~18 GFLOPS per core and that this unsurprisingly scales to more than 40 TFLOPS while using 2800 workers. To measure remote I/O throughput we benchmarked the read, write bandwidth to S3 and our benchmarks show that we can get on average 30 MB/s write and 40 MB/s read per Lambda and that this also scales to more than 60 GB/s write and 80 GB/s read. Assuming that 16 such Lambdas are as powerful as a single server, we find that the performance from Lambda matches the S3 performance shown in Table~\ref{tab:ssd_vs_s3}. To measure the overheads for small updates, we also benchmarked 128-byte synchronous put/gets to two \texttt{c3.8xlarge} instances running in-memory Redis. We match the performance reported in prior benchmarks~\cite{redis-benchmarks} and get less than 1ms latency up to 1000 workers. 


\textbf{Applications}: In our research group we have had students use PyWren for applications as diverse as computational imaging, scientific instrument design, solar physics, and object recognition. Working with heliphysicists at NASA's Solar Dynamics Observatory, we have
used PyWren for extracting relevant features across 16TB of solar imaging data for solar
flare prediction. 
Working with applied physics colleagues, we have used PyWren to design novel types
of microscope point-spread functions for 3d superresolution microscopy. This necessitates 
rapid and repeat evaluation of a complex physics-based optical model inside an inner loop.

\subsection{Generality for the rest of us ?}
\label{sec:apps}
While the map primitive in PyWren covers a number of applications, it prohibits any coordination among the various tasks. We next look at how stateless functions along with high performance storage can also be used as a flexible building block to develop more complex abstractions. 



\vspace{0.01in}
\textbf{Map + monolithic Reduce}
The first abstraction we consider is one where output from all the map operations is collected on to one machine (similar to \texttt{gather} in HPC literature) for further processing. 
We find this pattern covers a number of classical machine learning workloads which consist of a featurization (or ETL) stage that converts large input data into features and then a learning stage where the model is built using SVMs or linear classifiers. In such workloads, the featurization requires parallel processing but the generated features are often small and fit on a single large machine~\cite{canny2013big}. These applications can be implemented using a \emph{map} that runs using stateless functions followed by a learning stage that runs on a single multi-core server using efficient multi-core libraries~\cite{recht2011hogwild}. The wide array of machine choices in the cloud means that this approach can handle learning problems with features up to 2TB in size~\cite{x1-instance}. 

\begin{table}[!t]
\begin{center}
\begin{tabular}{c|cc}
\hline
 phase & mean & std \\
\hline
lambda start latency & 9.7s & 29.1s \\
lambda setup time & 14.2s & 5.2s \\
featurization & 112.9s & 10.2s \\
result fetch & 22.0s & 10.0s \\
fit linear classifier & 4.3s & 0.5s \\
\end{tabular}
\vspace{-0.0in}
\ncaption{\bf Time taken for featurization and classification}
\label{tab:map_mono_reduce}
\end{center}
\vspace{-0.3in}
\end{table}
As an example application we took off-the-shelf image featurization code \cite{douze2009evaluation} and performed cropping, scaling, and GIST image featurization~\cite{oliva2001modeling} of the 1.28M images in the ImageNet LargeScale Visual Recognition Challenge~\cite{ILSVRC15}. We run the end-to-end featurization using 3000 workers on AWS Lambda. and store the features on S3. This takes 113 seconds and following that we run a monolithic \emph{reduce} on a single \texttt{r4.16xlarge} instance. Fetching the features from S3 to this instance only takes 22s and building a linear classifier using NumPy and Intel MKL libraries takes 4.3s. Thus, we see that this model is a good fit where a high degree of parallelism is initially required to do ETL / featurization but a single node is sufficient (and most efficient~\cite{mcsherry2015scalability}) for model building.

\textbf{MapReduce}: For more general purpose coordination, a commonly used programming model is the bulk-synchronous processing (BSP) model. To implement the BSP model, in addition to parallel task execution, we need to perform data shuffles across stages. The availability of high-bandwidth remote storage provides an natural mechanism to implement such shuffles. Using S3 to store shuffle data, we implemented a word count program in PyWren. On the Amazon reviews~\cite{mcauley2015image} dataset consisting of 83.68M product reviews split across 333 partitions, this program took 98.6s. We ran a similar program using PySpark. Using 85 \texttt{r3.xlarge} instances, each having 4 cores to match the parallelism we had with PyWren, the Spark job took 84s. The slow down is from the lack of parallel shuffle block reads in PyWren and some stragglers while writing/reading from S3. Despite that we see that PyWren is only around 17\% slower than Spark and our timings do not include the 5-10 minutes it takes to start the Spark instances.

We also run the Daytona sort benchmark~\cite{sortbenchmark} on 1TB input, to see how PyWren handles a shuffle-intensive workload. We implemented the Terasort~\cite{terasort} algorithm to perform sort in two stages: a partition stage that range-partitions the input and writes out to intermediate storage, and a merge stage that, for each partition, merges and sorts all intermediate data for that partition and writes out the sorted output. Due to the resource limitation on each Lambda worker, we need at least 2500 tasks for each stage. This results in $2500^{2}$, or 6,250,000 intermediate files to shuffle in between. While S3 does provide abundant I/O bandwidth to Lambda, it falls short on sustaining high request throughput. Therefore, we use S3 only for storing input and writing output and deploy a Redis cluster (with \texttt{cache.m4.10xlarge} nodes) for intermediate storage. Figure~\ref{fig:sort_time} shows the end-to-end performance with varying numbers of concurrent Lambda workers and Redis shards, with breakdown of task time. We see that higher level of parallelism does greatly improve job performance (up to 500 workers) until Redis throughput becomes a bottleneck. From 500 to 1000 workers, the Redis I/O time increases by 42\%. Fully leveraging this parallelism requires more Redis shards, as shown by the 44\% improvement with 30 shards. Interestingly, adding more resources does not necessarily increase total cost due to the reduction in latency with scale (Figure~\ref{fig:sort_cost}).\footnote{Lambda bills in 100ms increments. Redis is charged per hour and is prorated here to seconds per CloudSort benchmark rules~\cite{sortbenchmark}.} Supporting a larger sort, e.g., 100TB , does become quite challenging, as the number of intermediate files increases quadratically. We plan to investigate more efficient solutions.


\textbf{Parameter Servers}: Finally using low-latency, high throughput key-value stores like Redis, RAMCloud~\cite{low-latency} we can also implement parameter-server~\cite{abadi2016tensorflow, li2014scaling} style applications in PyWren. For example, we can implement \textsc{Hogwild!} stochastic gradient descent by having each function compute the gradients based on the latest version of shared model. Since the only coordination across functions happens through the parameter server, such applications fit very well into the stateless function model. Further we can use existing support for server-side scripting~\cite{redislua} in key value stores to implement features like range updates and flexible consistency models~\cite{li2014scaling}. However, currently this model is not easy to use as unlike S3, the ElasticCache service requires users to select a cache server type and capacity. 








\section{Discussion}
While we studied the performance provided by existing infrastructure in the previous section, there are a number of systems aspects that need to be addressed to enable high performance data processing. 


\textbf{Resource balance}: One of the primary challenges in a serverless design is in how a function's resource usage is allocated and as we mentioned in~\xref{sec:pywren}, the existing limits are quite low. The fact that the functions are stateless and need to transfer both input and output over the network can help cloud providers come up with some natural heuristics. For example if we consider the current constraints of AWS Lambda we see that each Lambda has around 35 MB/s bandwidth to S3 and can thus fill up its memory of 1.5GB in around 40s. Assuming it takes 40s to write output, we can see that the running time of 300s is appropriately proportioned for around 80s of I/O and 220s of compute. As memory capacity and network bandwidths grow, this rule can be used to automatically determine memory capacity given a target running time. 

\textbf{Pricing} At the time of writing Lambda is priced at $\sim$\$0.06 per GB-hour of execution, measured in 100ms-increments. Lambda is thus only $\sim$2$\times$ more expensive than on-demand instances. This cost premium seems worthwhile given substantially finer-grained billing, much greater elasticity, and the fact that many dedicated clusters are often running at 50\% utilization. Another benefit that stems from PyWren's disaggregated architecture is that cost estimation or even cost prediction becomes much simpler. In the future we plan to explore techniques that can automatically predict the cost of a computation.


\textbf{Scalable Scheduling}: A number of cluster scheduling papers~\cite{wilkes2013omega,ousterhout2013sparrow,ousterhout2013case,isard2009quincy} have looked at providing low latency scheduling for data parallel frameworks running on servers. However, to implement such scheduling frameworks on top of stateless functions, we need to handle the fact that information about the cluster status (i.e., which containers are free, input locations) is only available to the infrastructure provider, while the structure of the job (i.e. how functions depend on each other) is only available to the user. 
In the future we plan to study what information needs to exposed by cloud providers and if scheduling techniques like offers~\cite{hindman2011mesos} can handle this separation. 

\textbf{Distributed Storage}: With the separation of storage and compute in the PyWren programming model, a number of performance challenges translate into the need for more efficient distributed storage systems. Our benchmarks in~\xref{sec:pywren} showed the limitations of current systems, especially for supporting large shuffle-intensive workloads, and we plan to study how we can enable a flat-datacenter storage system in terms of latency and bandwidth~\cite{fds}. Further, our existing benchmarks also show the limitation of not lacking API support for append in systems like S3 and we plan to develop a common API for storage backends that power serverless computation. 



\textbf{Launch Overheads}: Finally one of the main drawbacks in our current implementation is that function invocation can take up to 20-30 seconds ($\sim$10\% of the execution time) without any caching. This is partly due to lambda invocation rate limits imposed by AWS and partly due to the time taken to setup our custom python runtime. We plan to study if techniques used to make VM forks cheaper~\cite{snowflock}, like caching containers or layering filesystems can be used to improve latency. We also plan to see if the scheduler can be modified to queue functions before their inputs are ready to handle launch overheads.


\textbf{Other applications}: While we discussed data analytics applications that fit well with the serverless model, there are some applications that do not fit today. Applications that use specialized hardware like GPUs or FPGAs are not supported by AWS Lambda, but we envision that more general hardware support will be available in the future. However, for applications like particle simulations, which require a lot of coordination between long running processes, the PyWren model of using stateless functions with remote storage might not be a good fit. Finally, while we primarily focused on existing analytics applications in this paper, the serverless model has also been used successfully in other domains like video compression~\cite{fouladi2017encoding}. 


\section{Conclusion}
The server-oriented focus of existing data processing systems in the cloud presents a high barrier for a number of users.
In this paper we propose that using stateless functions with remote storage, we can build a data processing system that inherits the elasticity, simplicity of the serverless model while providing a flexible building block for more complex abstractions.





{
\bibliographystyle{acm}
\bibliography{paper}

\begin{thebibliography}{10}

\bibitem{abadi2016tensorflow}
{\sc Abadi, M., Barham, P., Chen, J., Chen, Z., Davis, A., Dean, J., Devin, M.,
  Ghemawat, S., Irving, G., Isard, M., et~al.}
\newblock Tensorflow: A system for large-scale machine learning.
\newblock In {\em OSDI\/} (2016).

\bibitem{armbrust2010view}
{\sc Armbrust, M., Fox, A., Griffith, R., Joseph, A.~D., Katz, R., Konwinski,
  A., Lee, G., Patterson, D., Rabkin, A., Stoica, I., et~al.}
\newblock A view of cloud computing.
\newblock {\em CACM 53}, 4 (2010), 50--58.

\bibitem{asanovic2014firebox}
{\sc Asanovic, K., and Patterson, D.}
\newblock Firebox: A hardware building block for 2020 warehouse-scale
  computers.
\newblock In {\em FAST\/} (2014).

\bibitem{aws-lambda-mapred}
{Serverless Reference Architecture: MapReduce}.
\newblock \url{https://github.com/awslabs/lambda-refarch-mapreduce}.

\bibitem{canny2013big}
{\sc Canny, J., and Zhao, H.}
\newblock Big data analytics with small footprint: Squaring the cloud.
\newblock In {\em KDD\/} (2013).

\bibitem{carriero1998linda}
{\sc Carriero, N., and Gelernter, D.}
\newblock Linda in context.
\newblock {\em CACM 32}, 4 (Apr. 1989).

\bibitem{cloudpickle}
cloudpickle: Extended pickling support for python objects.
\newblock \url{https://github.com/cloudpipe/cloudpickle}.

\bibitem{douze2009evaluation}
{\sc Douze, M., J{\'e}gou, H., Sandhawalia, H., Amsaleg, L., and Schmid, C.}
\newblock Evaluation of gist descriptors for web-scale image search.
\newblock In {\em ACM International Conference on Image and Video Retrieval\/}
  (2009).

\bibitem{ethernet-standards}
{IEEE P802.3ba, 40Gb/s and 100Gb/s Ethernet Task Force}.
\newblock \url{http://www.ieee802.org/3/ba/}.

\bibitem{fang2015interruptible}
{\sc Fang, L., Nguyen, K., Xu, G., Demsky, B., and Lu, S.}
\newblock {Interruptible tasks: Treating memory pressure as interrupts for
  highly scalable data-parallel programs}.
\newblock In {\em SOSP\/} (2015).

\bibitem{fouladi2017encoding}
{\sc Fouladi, S., Wahby, R.~S., Shacklett, B., Balasubramaniam, K.~V., Zeng,
  W., Bhalerao, R., Sivaraman, A., Porter, G., and Winstein, K.}
\newblock {Encoding, Fast and Slow: Low-Latency Video Processing Using
  Thousands of Tiny Threads}.
\newblock In {\em NSDI\/} (2017).

\bibitem{diskirrelevant}
{\sc {G. Ananthanarayanan, A. Ghodsi, S. Shenker, I. Stoica}}.
\newblock {Disk-Locality in Datacenter Computing Considered Irrelevant}.
\newblock In {\em Proc. HotOS\/} (2011).

\bibitem{gao2016network}
{\sc Gao, P.~X., Narayan, A., Karandikar, S., Carreira, J., Han, S., Agarwal,
  R., Ratnasamy, S., and Shenker, S.}
\newblock Network requirements for resource disaggregation.
\newblock In {\em OSDI\/} (2016).

\bibitem{han2013network}
{\sc Han, S., Egi, N., Panda, A., Ratnasamy, S., Shi, G., and Shenker, S.}
\newblock Network support for resource disaggregation in next-generation
  datacenters.
\newblock In {\em HotNets\/} (2013).

\bibitem{han2013large}
{\sc Han, S., and Ratnasamy, S.}
\newblock Large-scale computation not at the cost of expressiveness.
\newblock In {\em HotOS\/} (2013).

\bibitem{hendrickson2016serverless}
{\sc Hendrickson, S., Sturdevant, S., Harter, T., Venkataramani, V.,
  Arpaci-Dusseau, A.~C., and Arpaci-Dusseau, R.~H.}
\newblock {Serverless computation with OpenLambda}.
\newblock In {\em HotCloud\/} (2016).

\bibitem{herodotou2011starfish}
{\sc Herodotou, H., Lim, H., Luo, G., Borisov, N., Dong, L., Cetin, F.~B., and
  Babu, S.}
\newblock Starfish: A self-tuning system for big data analytics.
\newblock In {\em CIDR\/} (2011).

\bibitem{hettrick_2014_14809}
{\sc Hettrick, S., Antonioletti, M., Carr, L., Chue~Hong, N., Crouch, S.,
  De~Roure, D., Emsley, I., Goble, C., Hay, A., Inupakutika, D., Jackson, M.,
  Nenadic, A., Parkinson, T., Parsons, M.~I., Pawlik, A., Peru, G., Proeme, A.,
  Robinson, J., and Sufi, S.}
\newblock Uk research software survey 2014.
\newblock \url{https://doi.org/10.5281/zenodo.14809}, Dec. 2014.

\bibitem{hindman2011mesos}
{\sc Hindman, B., Konwinski, A., Zaharia, M., Ghodsi, A., Joseph, A., Katz, R.,
  Shenker, S., and Stoica, I.}
\newblock {Mesos: A Platform for Fine-Grained Resource Sharing in the Data
  Center}.
\newblock In {\em Proc. NSDI\/} (2011).

\bibitem{hp-machine}
{HP The Machine: Our vision for the Future of Computing}.
\newblock \url{https://www.labs.hpe.com/the-machine}.

\bibitem{isard2009quincy}
{\sc Isard, M., Prabhakaran, V., Currey, J., Wieder, U., Talwar, K., and
  Goldberg, A.}
\newblock {Quincy: Fair Scheduling for Distributed Computing Clusters}.
\newblock In {\em Proc. SOSP\/} (2009), pp.~261--276.

\bibitem{snowflock}
{\sc Lagar-Cavilla, H.~A., Whitney, J.~A., Scannell, A.~M., Patchin, P.,
  Rumble, S.~M., de~Lara, E., Brudno, M., and Satyanarayanan, M.}
\newblock Snowflock: Rapid virtual machine cloning for cloud computing.
\newblock In {\em EuroSys\/} (2009).

\bibitem{li2014scaling}
{\sc Li, M., Andersen, D.~G., Park, J.~W., Smola, A.~J., Ahmed, A., Josifovski,
  V., Long, J., Shekita, E.~J., and Su, B.-Y.}
\newblock Scaling distributed machine learning with the parameter server.
\newblock In {\em OSDI\/} (2014).

\bibitem{mcauley2015image}
{\sc McAuley, J., Targett, C., Shi, Q., and Van Den~Hengel, A.}
\newblock Image-based recommendations on styles and substitutes.
\newblock In {\em SIGIR\/} (2015).

\bibitem{mcsherry2015scalability}
{\sc McSherry, F., Isard, M., and Murray, D.~G.}
\newblock {Scalability! but at what COST?}
\newblock In {\em HotOS\/} (2015).

\bibitem{momcheva2015software}
{\sc Momcheva, I., and Tollerud, E.}
\newblock {Software Use in Astronomy: an Informal Survey}.
\newblock {\em arXiv 1507.03989\/} (2015).

\bibitem{fds}
{\sc Nightingale, E.~B., Elson, J., Fan, J., Hofmann, O., Howell, J., and
  Suzue, Y.}
\newblock Flat datacenter storage.
\newblock In {\em OSDI\/} (2012).

\bibitem{recht2011hogwild}
{\sc Niu, F., Recht, B., Re, C., and Wright, S.}
\newblock Hogwild: A lock-free approach to parallelizing stochastic gradient
  descent.
\newblock In {\em NIPS\/} (2011).

\bibitem{oliva2001modeling}
{\sc Oliva, A., and Torralba, A.}
\newblock Modeling the shape of the scene: A holistic representation of the
  spatial envelope.
\newblock {\em International Journal of computer vision 42}, 3 (2001),
  145--175.

\bibitem{openwhisk}
{OpenWhisk}.
\newblock \url{https://developer.ibm.com/openwhisk/}.

\bibitem{ousterhout2013case}
{\sc Ousterhout, K., Panda, A., Rosen, J., Venkataraman, S., Xin, R.,
  Ratnasamy, S., Shenker, S., and Stoica, I.}
\newblock The case for tiny tasks in compute clusters.
\newblock In {\em HotOS\/} (2013).

\bibitem{ousterhout2013sparrow}
{\sc Ousterhout, K., Wendell, P., Zaharia, M., and Stoica, I.}
\newblock Sparrow: distributed, low latency scheduling.
\newblock In {\em SOSP\/} (2013).

\bibitem{terasort}
{\sc O’Malley, O.}
\newblock {TeraByte Sort on Apache Hadoop}.
\newblock \url{http://sortbenchmark.org/YahooHadoop.pdf}.

\bibitem{peng2010large}
{\sc Peng, D., and Dabek, F.}
\newblock Large-scale incremental processing using distributed transactions and
  notifications.
\newblock In {\em OSDI\/} (2010).

\bibitem{power2010piccolo}
{\sc Power, R., and Li, J.}
\newblock Piccolo: Building fast, distributed programs with partitioned tables.
\newblock In {\em OSDI\/} (2010).

\bibitem{redislua}
Redis server side scripting.
\newblock \url{https://redis.io/commands/eval}.

\bibitem{redis-benchmarks}
Redis benchmarks.
\newblock \url{https://redis.io/topics/benchmarks}.

\bibitem{low-latency}
{\sc Rumble, S.~M., Ongaro, D., Stutsman, R., Rosenblum, M., and Ousterhout,
  J.~K.}
\newblock {It's Time for Low Latency}.
\newblock In {\em Proc. HotOS\/} (2011).

\bibitem{ILSVRC15}
{\sc Russakovsky, O., Deng, J., Su, H., Krause, J., Satheesh, S., Ma, S.,
  Huang, Z., Karpathy, A., Khosla, A., Bernstein, M., Berg, A.~C., and Fei-Fei,
  L.}
\newblock {ImageNet Large Scale Visual Recognition Challenge}.
\newblock {\em International Journal of Computer Vision (IJCV) 115}, 3 (2015),
  211--252.

\bibitem{wilkes2013omega}
{\sc Schwarzkopf, M., Konwinski, A., Abd-El-Malek, M., and Wilkes, J.}
\newblock {Omega: flexible, scalable schedulers for large compute clusters}.
\newblock In {\em Proc. EuroSys\/} (2013).

\bibitem{colin-trends}
{\sc Scott, C.}
\newblock Latency trends.
\newblock \url{http://colin-scott.github.io/blog/2012/12/24/latency-trends/}.

\bibitem{shvachko2010hadoop}
{\sc Shvachko, K., Kuang, H., Radia, S., and Chansler, R.}
\newblock {The Hadoop Distributed File System}.
\newblock In {\em Mass storage systems and technologies (MSST)\/} (2010).

\bibitem{sortbenchmark}
{Sort Benchmark}.
\newblock \url{http://sortbenchmark.org}.

\bibitem{spark-gc-blog}
{Tuning Java Garbage Collection for Apache Spark Applications}.
\newblock \url{https://goo.gl/SIWlqx}.

\bibitem{spark-gc-tuning}
{Tuning Spark}.
\newblock
  \url{https://spark.apache.org/docs/latest/tuning.html\#garbage-collection-tuning}.

\bibitem{yarn}
{\sc Vavilapalli, V.~K., Murthy, A.~C., Douglas, C., Agarwal, S., Konar, M.,
  Evans, R., Graves, T., Lowe, J., Shah, H., Seth, S., et~al.}
\newblock {Apache Hadoop YARN: Yet another resource negotiator}.
\newblock In {\em SoCC\/} (2013).

\bibitem{venkataraman2016ernest}
{\sc Venkataraman, S., Yang, Z., Franklin, M., Recht, B., and Stoica, I.}
\newblock Ernest: Efficient performance prediction for large-scale advanced
  analytics.
\newblock In {\em NSDI\/} (2016).

\bibitem{x1-instance}
{X1 instances}.
\newblock \url{https://aws.amazon.com/ec2/instance-types/x1/}.

\bibitem{zaharia2010spark}
{\sc Zaharia, M., Chowdhury, M., Das, T., Dave, A., Ma, J., McCauley, M.,
  Franklin, M., Shenker, S., and Stoica, I.}
\newblock {Resilient Distributed Datasets: A Fault-Tolerant Abstraction for
  In-Memory Cluster Computing}.
\newblock In {\em Proc. NSDI\/} (2011).

\end{thebibliography}
}

\end{document}